\documentclass[twocolumn,a4paper,accepted=2025-05-02]{quantumarticle}
\pdfoutput=1

\usepackage[utf8]{inputenc}
\usepackage[english]{babel}
\usepackage[T1]{fontenc}
\usepackage{amsmath}
\usepackage{hyperref}

\usepackage{amssymb,amstext}
\usepackage{color}
\usepackage{tikz}
\usepackage{graphicx}

\usepackage{xcolor}

\newcommand{\racionalni}{\ensuremath{\mathbb{Q}}}
\newcommand{\realni}{\ensuremath{\mathbb{R}}}

\newcommand{\meas}{\mathop{\rm meas}\nolimits}
\newcommand{\mycirc}{\mathop{\circ}\nolimits}
\newcommand{\Diff}{Di\!f\!\!f}
\newcommand{\ds}{\displaystyle}

\newcommand{\cC}{{\cal C}}

\newcommand{\cM}{{\cal M}}

\newcommand{\cO}{{\cal O}}
\newcommand{\cP}{{\cal P}}

\newcommand{\cR}{{\cal R}}

\newtheorem{Theorem}{Theorem}
\newtheorem{conjecture}[Theorem]{Conjecture}

\begin{document}

\title{Operational verification of the existence of a spacetime manifold}

\author{Nikola Paunkovi\'c}
\affiliation{Instituto de Telecomunica\~coes and Departamento de Matem\'atica, Instituto Superior T\'ecnico, Universidade de Lisboa, Avenida Rovisco Pais 1, 1049-001, Lisboa, Portugal}
\email{npaunkov@math.tecnico.ulisboa.pt}
\orcid{0000-0002-9345-4321}

\author{Marko Vojinovi\'c}
\affiliation{Institute of Physics, University of Belgrade, Pregrevica 118, 11080 Belgrade, Serbia}
\email{vmarko@ipb.ac.rs}
\orcid{0000-0001-6977-4870}


\begin{abstract}
We argue that there exists an operational way to establish the observability of the notions of space and time. Specifically, we propose a theory-independent protocol for a gedanken-experiment, whose outcome is a signal establishing the observability of the spacetime manifold, without a priori assuming its existence. The experimental signal contains the information about the dimension and the topology of spacetime (with the currently achievable precision), and establishes its manifold structure, while respecting its underlying diffeomorphism symmetry. We also introduce and discuss appropriate criteria for the concept of emergence of spacetime, which any tentative theoretical model of physics must satisfy in order to claim that spacetime does emerge from some more fundamental concepts.
\end{abstract}

\maketitle

\section{Introduction}

Both in physics and philosophy, one often asks deep fundamental questions such as ``What is time?'', ``What is space?'', ``Do they objectively exist or not?'', and similar. While a lot has been said and written about these questions throughout history of science and philosophy \cite{Huggett1999,Dainton2001,Alexander1956,Vailati1997,Sklar1992,Carnap2012,Jammer1954,Wheeler1990}, there appear to be very few attempts to address these questions in terms of {\em experimental evidence} for the notions of time and space. On one hand, we all have intuitive feeling for both, since we heavily rely on them in daily life, and this intuition is partially based on some experimental evidence. On the other hand, conceptual analysis of the {\em objective existence} of space and time turned out to be quite a hard problem, due to the high level of symmetry properties of space and time, encoded in the principle of general relativity.

Namely, there is a long-standing common claim that the points on a spacetime manifold are physically unobservable, given the principle of general relativity, i.e., because we expect all physics to be diffeomorphism-invariant and background-independent. As far as the statement goes, one cannot distingush between ``this point'' versus ``that point'' of spacetime itself, but only ``the point where fields have this value'' versus ``the point where fields have that value'', along the lines of the paradigm of relational approach to physics (see for example \cite{Rovelli2004}).

Loosely speaking, the basic argument for the unobservability of individual spacetime points goes as follows. If we choose one point of spacetime (by specifying its coordinates in some coordinate system), determine the values of all fields at that point, and then perform an ``active diffeomorphism'' (permutation of manifold points), we ``move'' all physics from that point to another point. After that, we can perform a ``passive diffeomorphism'' (choice of a different manifold chart), to undo the active one, i.e., we use the same set of numbers as coordinates for the new point in the new coordinate chart as we have used for the old point in old coordinates. Given that physics does not change throughout the process, we conclude that one cannot distinguish between the ``old spacetime point'' and the ``new spacetime point''. Thus, individual spacetime points are unobservable.

While this is correct in itself, apparently there are proposals that go even further, and generalize this argument to deny the existence of the spacetime manifold altogether. The argument could possibly be paraphrased in the following form --- if spacetime points are not observable, they do not objectively exist, so therefore the manifold itself does not objectively exist.

The purpose of this work is to challenge this generalization. Namely, our statement is the following: the fact that we cannot observe individual points does \textit{not} imply that the spacetime manifold as a whole cannot be observed, or that it does not objectively exist in nature. In particular, a manifold has properties which are invariant with respect to diffeomorphisms --- specifically, its dimension and its topology, and a priori both of these might in principle be observable. The main point of this work is to demonstrate that these properties of a manifold {\em indeed are observable}. As a consequence of this, we argue that the \textit{spacetime manifold is itself observable}, without violating either the diffeomorphism symmetry or background independence. Additionally, our argument for the observability of spacetime, presented below, relies on an operational, theory-independent experimental protocol, incorporated in a specific proposed gedanken-experiment. It is thus mainly based on experimental evidence, rather than on some kind of theoretical or metaphysical assumptions.

In order to further emphasise this last point, one can imagine a hypothetical scenario involving an artificial inteligence (AI) implemented within a memory of a computer, without having any a priori notion about time and space. One can further imagine that this AI is perfectly capable of performing complex mathematical analyses. In such a setup, the experimenter could perform our proposed protocol, and feed AI the resulting experimental data to analyse it. The outcome of the analysis performed by the AI would then be a conclusion that the experimenter ``lives in a space and time'' of dimension 4 and simply connected topology, despite the fact that AI itself does not have any intuitive or a priori notions of either space or time. This hypothetical scenario emphasises the independence of our proposed experimental protocol on any a priori notions of time and space that an expermenter may be subject to, because the AI would reach the same conclusion without such a priori notions.

The layout of the paper is as follows. Sections \ref{SecMechanics} and \ref{SecFieldTheory} discuss the details of the thought experiment in mechanics and field theory, respectively. The two cases have been separated into two sections merely for the pedagogical purpose of gradually introducing the relevant analysis and techniques, while in principle the analysis in the context of mechanics is merely a special case of the analysis in the context of field theory. Section \ref{SecConclusions} contains a discussion of various related aspects of the analysis, as well as a discussion of several topics which put our results in a wider context. Among other topics, we discuss the notion of spacetime emergence. In particular, we provide nontrivial criteria that arguably have not been satisfied by any of the existing proposals aiming to describe the emergence of spacetime. Appendix \ref{AppA} contains some important technical results needed for the analysis.

\section{\label{SecMechanics}Mechanics and time}

In order to have a clear understanding of the ideas proposed in this work, it is prudent to first discuss the toy-example of a $(0+1)$-dimensional spacetime manifold, i.e., the time manifold. Various key properties of the analysis can be discussed using simple mechanical systems, such as pendulums, while the proper $(3+1)$-dimensional spacetime manifold is postponed for Section \ref{SecFieldTheory}.

Of course, throughout the text we uphold an initial assumption that the concepts of time and spacetime are not observable a priori.

\subsection{System with one observable}

Let us introduce the following gedanken-experiment.  We are given a swinging pendulum, denoted $A$. Without any assumptions about Newtonian mechanics (since it relies on the a priori notion of time), our objective is to describe the motion of the pendulum, as precisely as we can. The measurement apparatus we have available is a camera that can take still photos of the pendulum, along with a ruler that can measure the distances on the photos.

Start by taking $N$ photos of the pendulum, completely randomly, and use a ruler to measure the signed distance $a_k$ of the pendulum from its vertical axis, for every photograph $k \in \{ 1,\dots,N \}$. The distance is signed in the sense that if the pendulum is left of its axis we consider the distance to be negative, while if the pendulum is right of its axis, the sign of distance is positive (the choices of left and right are merely a convention and do not influence any conclusions). Moreover, we want to eliminate any information about the ``time order'' in which the photos might have been taken, so we remember only the following {\em unordered} set of measurements, describing what we can tell about the motion of pendulum $A$,
\begin{equation}
A = \{ a_1, a_2, \dots, a_N \}\,,
\end{equation}
where
\begin{equation}
a_k \in S_A \equiv [a_{min},a_{max}] \subset \realni\,, \quad \forall k = 1,\dots,N\,.
\end{equation}
Here $a_{min}$ and $a_{max}$ denote the left and right amplitudes of the pendulum, respectively. In usual circumstances one should obtain $a_{min} = - a_{max}$, but for our purposes this equality does not really matter. Also, we assume that no two measurements produce exactly equal values of $a$, so that $a_i \neq a_j$ for all measurements $i$ and $j$. This assumption is just for convenience, and we will discuss later the conceptual case in which the measurement results are repeating themselves, and always falling into a discrete set of outcomes.

\begin{figure}[ht]
\begin{center}
\includegraphics[scale=0.21]{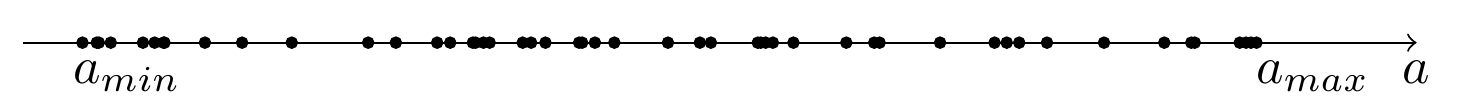}
\end{center}
\caption{\label{figure1}A scatter plot for a single pendulum system.}
\end{figure}

Given the measurement dataset $A$, one can draw it as a one-dimensional scatter plot on the $a$-axis, which looks similar to the plot in Figure \ref{figure1}.
If we take additional $N$ measurements (i.e., a total of $2N$), the scatter plot will look similar, only with twice as many points in the segment $S_A$. In the limit $N\to \infty$, the scatter plot will become dense, and practically fill the whole segment $S_A$, given that we can measure distances $a_k$ with an arbitrarily small but still nonzero precision.  The notion of arbitrary small distance is not in conflict with quantum mechanics, since we are measuring only the position of the pendulum, and not its momentum. With sufficiently many measurements, we can describe the motion of the pendulum using the whole set $S_A$ of possible positions it might be in, and we can call this set the \textit{configuration space} of the pendulum.

However, our naive intuition suggests that such a description of pendulum's motion is not completely satisfactory, since each photo demonstrates that the pendulum is at some particular distance $a$, while the configuration space of the pendulum describes the pendulum merely ``in all positions'', suggesting that we are missing some extra information. We therefore ask for a more precise description of the pendulum's motion. The precise formalization of this ``intuition'' will be given below, but at this stage let us appeal to the ideas of relationalism, and try to obtain the missing information by {\em comparing} the motion of the pendulum against {\em some other physical system}.

\subsection{System with two observables}

\begin{figure}[ht]
\begin{center}
\includegraphics[scale=0.26]{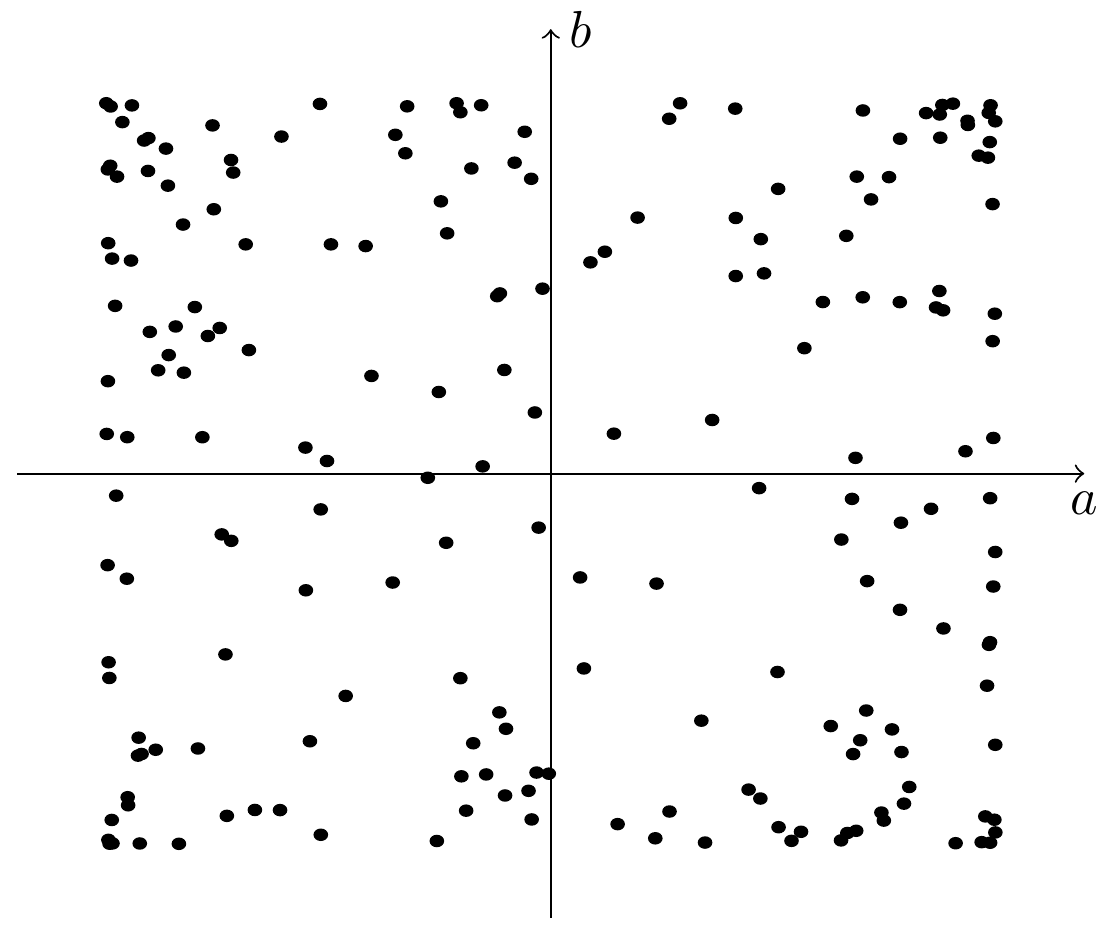}
\end{center}
\caption{\label{figure2}An ordinary scatter plot for a double pendulum system.}
\end{figure}

\begin{figure}[ht]
\begin{center}
\includegraphics[scale=0.26]{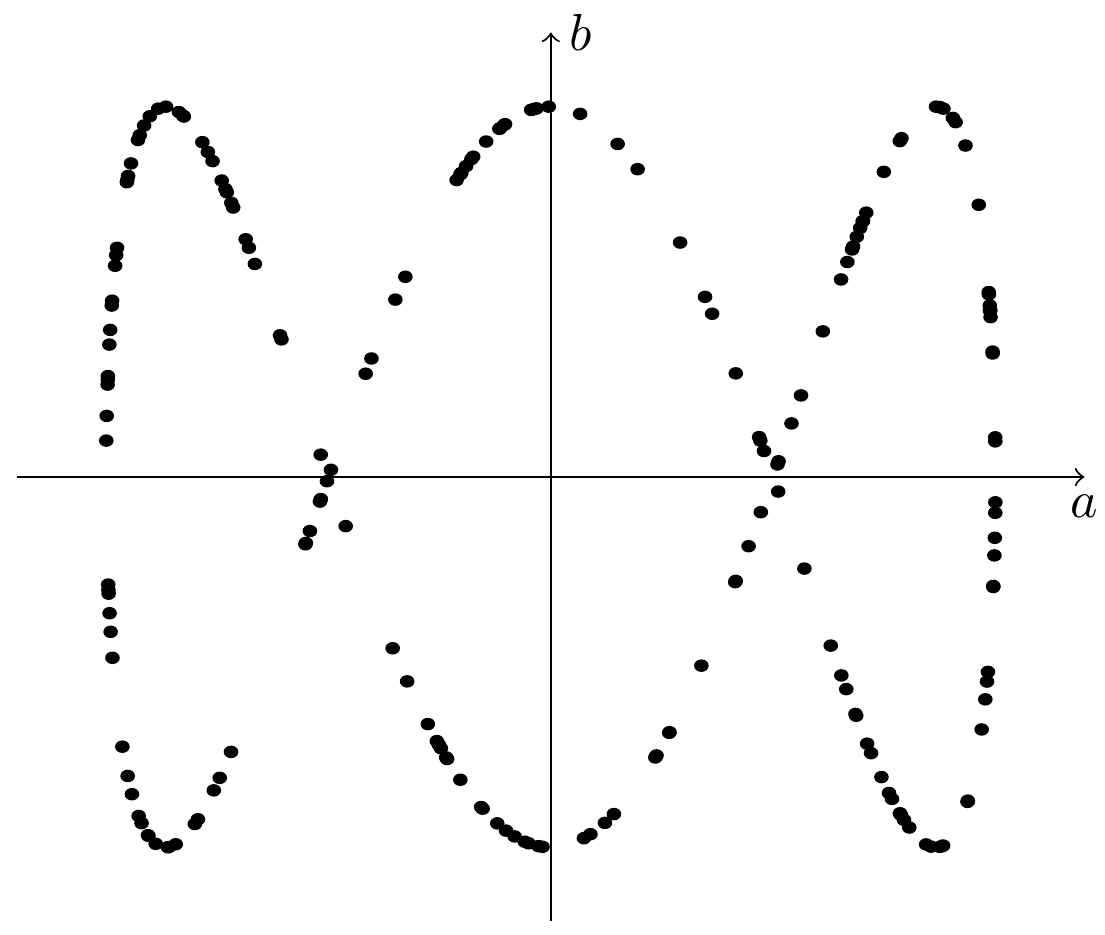}
\end{center}
\caption{\label{figure3}An extraordinary scatter plot for a double pendulum system.}
\end{figure}

For lack of a better idea, let us introduce another pendulum, denoted $B$, in addition to $A$. In a generic situation, the pendulum $B$ may have different length and other properties compared to $A$ (we discuss this and other properties of the protocol setup in more detail in Subsection \ref{SubSec:InterpretingTimeMfld}). We put the two pendulums side by side, and randomly take photos of the whole system. We then cut each photo into two pieces, such that each pendulum is displayed separately on its half of the photo. As before, we now shuffle all photos to erase the order in which the photos were taken, as well as the relation which photo of $A$ is paired to which photo of $B$. We end up with the following {\em unordered} datasets for the two pendulums:
\begin{equation} \label{TwoPendulumDataset}
A = \{ a_1, a_2, \dots, a_N \}\,, \quad B = \{ b_1, b_2, \dots, b_N \}\,.
\end{equation}
After sufficiently many measurements, we can establish the individual configuration spaces for each pendulum like before,
\begin{equation}
S_A \equiv [a_{min},a_{max}] \subset \realni\,, \quad S_B \equiv [b_{min},b_{max}] \subset \realni\,.
\end{equation}
Again, we have an intuition that this is not a complete description of the motion of the two pendulums. But now we can formalize this intuition as follows. Given the individual configuration spaces $S_A$ and $S_B$, a priori one can say that the \textit{joint} configuration space will be the set $S_A\times S_B$. One can visualize this experimental result within this set graphically, as follows. Pick a random permutation $\pi$ of $N$ elements, and construct the following $2\times N$ matrix,
\begin{equation}
\left(
\begin{array}{cccc}
a_1 & a_2 & \dots & a_N \\
b_{\pi(1)} & b_{\pi(2)} & \dots & b_{\pi(N)} \\
\end{array}
\right)\,,
\end{equation}
which describes an arbitrary ``pairing'' of each measurement $a_k$ to some measurement $b_{\pi(k)}$. These pairings define $N$ points on a scatter plot in the space $S_A\times S_B$, which will typically look like the one in Figure \ref{figure2}.
One can draw a similar scatter plot for every choice of the permutation function $\pi$. Since for $N$ measurements there are $N!$ possible permutations, there are also $N!$ scatter plots. But, as an experimental result of our thought experiment, among all those scatter plots, there will be \textit{one special plot}, corresponding to the permutation denoted as $\tilde{\pi}$, which is visually very distinguishable from the rest, and looks like the one in Figure \ref{figure3}.
Looking at the plot, it is straightforward to interpret it as a set of points on a $1$-dimensional curve, denoted $\cC$. The experimental fact that the joint configuration space $S_{A\cup B}$ is equal to the curve $\cC$ has one important consequence --- it captures our intuition of a more precise description of the motion of pendulums $A$ and $B$. This is because the curve $\cC$ is equivalent to a functional dependence (a \textit{correlation}) between observables $a$ and $b$, of the general form
\begin{equation} \label{CorrelationBetweenAandB}
F(a,b) = 0\,.
\end{equation}
The equation \eqref{CorrelationBetweenAandB} actually defines the curve $\cC$ as a set of all pairs $(a,b) \in S_A \times S_B$ which satisfy \eqref{CorrelationBetweenAandB}. At this point we can spell out the formalization of our intuition for the results of the thought experiment, in the form of the following conjecture.

\begin{conjecture}
(definition of special correlations).
The joint configuration space $S_{A \cup B}$ is a \textit{strict subspace} of $S_A\times S_B$, which is \textit{of measure zero} compared to $S_A\times S_B$:
\begin{equation}
S_{A \cup B} \subset S_A\times S_B\,, \qquad \frac{\meas(S_{A\cup B})}{\meas(S_A\times S_B)} = 0\,.
\end{equation}
The measure over these sets is defined as an induced measure from the standard Euclidean metric over $\realni^2$.
\end{conjecture}

Note that, while visual distinguishability is appealing, it is a particular pattern-recognition trait of a human brain. In order to have a more formal definition of why this particular diagram is ``special'', without appealing to our eyes and brains, we can resort to the statistical analysis described in detail in the Appendix \ref{AppA}. This can be implemented as a computer algorithm operating on dataset \eqref{TwoPendulumDataset}, eliminating any reliance on visual inspection of scatter plots. Also, the formal statistical analysis has an advantage of being applicable to higher dimensions, i.e., beyond the $2$- and $3$-dimensional scatter plots, in contrast to visual inspection by a human. This property will become important in Section \ref{SecFieldTheory}.

The ``special'' permutation $\tilde{\pi}$, corresponding to the ``special'' plot above, has several crucial properties:
\begin{conjecture}
(properties of the special correlations).
\begin{itemize}
\item {\em Existence}. Given a completely random set of numbers $a_k$ and $b_k$, $k=1,\dots,N$, the permutation $\tilde{\pi}$ does not necessarily exist. This is actually one of the definitional properties for a sequence of numbers to be ``random'' --- absence of any noticable correlation. It should be stressed that the existence of $\tilde{\pi}$ is a specific property of the dataset \eqref{TwoPendulumDataset}, which {\em fails to be random}, as opposed to an arbitrary set of numbers not obtained by experimental measurements of the two pednulums. The existence of $\tilde{\pi}$ is therefore {\em an experimental signal}.
\item {\em Self-reinforcement}. Suppose that, after taking $N$ measurements \eqref{TwoPendulumDataset}, we continue to take additional $M$ measurements,
\begin{equation} \label{SecondTwoPendulumDataset}
\begin{array}{lcl}
  A' & = & \ds \{ a_{N+1}, a_{N+2}, \dots, a_{N+M} \}\,, \\
  B' & = & \ds \{ b_{N+1}, b_{N+2}, \dots, b_{N+M} \}\,. \vphantom{\ds\int} \\
\end{array}
\end{equation}
We can then perform the same analysis independently over the three datasets --- the original dataset $(A,B)$, the new dataset $(A',B')$, and their union dataset $(A\cup A', B\cup B')$, to arrive at the three ``special'' permutations $\tilde{\pi}$, $\tilde{\pi}'$ and $\tilde{\Pi}$, defined over $N$, $M$ and $N+M$ numbers respectively. Then, it is again {\em an experimental result} that the restriction of $\tilde{\Pi}$ to the first $N$ numbers will be equal to the original permutation $\tilde{\pi}$, while the restriction of $\tilde{\Pi}$ to the remaining $M$ numbers will be equal to the new permutation $\tilde{\pi}'$. Thus, we have a piecewise-defined equation
\begin{equation}
\tilde{\Pi}(k) = \left\{
\begin{array}{ll}
\tilde{\pi}(k)\,, & k = 1,\dots,N\,, \\
\tilde{\pi}'(k)\,, & k = N+1,\dots,N+M\,, \\
\end{array}
\right.\!\!\!\!\!\!
\end{equation}
which holds for all possible choices of $N$ and $M$. Pictorially, this property means that once we draw scatter plots corresponding to $\tilde{\pi}$ and $\tilde{\pi}'$, the data points will remain nicely aligned along the {\em same curve}. In other words, the correlation which represents our experimental signal {\em reinforces itself} when one adds additional data.

Note that self-reinforcement is an important property in situations where one is interested in discussing the discrete-to-continuum limit. In the context of spacetime, the discrete-to-continuum limit has been extensively studied in causal set theory. See for example \cite{Butterfield} for details.

\item {\em Dimensionality}. In the limit $N\to\infty$, one can see that there exists a special permutation $\tilde{\pi}$ such that the data points are arranged in a dense way along the whole curve $\cC$. In other words, the actual {\em joint configuration space} $\cC \equiv S_{A\cup B}$ for pendulums $A$ and $B$ is not only a subset of measure zero in the direct product $S_A\times S_B$ of two individual configuration spaces, but also a {\em $1$-dimensional subset}. This is also an experimental result. Namely, it could have happened that $S_{A\cup B}$ is some patch in $S_A\times S_B$ of nonzero area ($2$-dimensional), or a discrete set of points ($0$-dimensional), or any combination thereof. But these alternative scenarios may happen only for non-experimental datasets, while for experimental datasets it always turns out that $S_{A\cup B}$ is $1$-dimensional.
\item {\em Topology}. The statistical analysis from Appendix \ref{AppA} provides us not only with the dimension, but also with a convenient set of patches along the curve $\cC$, which define a {\em basis of open sets}, giving rise to an induced {\em topology} on $\cC$. In the example of the curve in the picture, the topology is that of a circle, but it can also be open curve in other examples, as we shall discuss in more detail below.
\end{itemize}
\end{conjecture}

Once the existence and properties of the $1$-dimensional curve $\cC$ have been established, one may be tempted to call it the ``time manifold''. However, there are several issues associated with that, discussed in the next Subsection.

\subsection{\label{SubSec:InterpretingTimeMfld}Interpreting the results as a time manifold}

At the operational level, our protocol makes use of a device which takes a photo of two pendulums. One immediate question that arises regarding this setup is whether this implicitly assumes some a priori notion of ``simultaneity'', or maybe even ``time''. Namely, taking a single photo of two pendulums may be considered equivalent to ``simultaneously'' taking two photos of two pendulums individually. However, this is not equivalent --- we deliberately take {\em one single} photo of a {\em composite system}, rather than {\em two separate} photos of its {\em subsystems}. This is because the requirement that the two separate photos be taken ``simultaneously'' would indeed necessarily assume an a priori notion of time. On the other hand, one can circumvent any such assumption by taking a single photo to {\em operationally define} the notion of ``simultaneity'', without a priori assuming time. Our definition is as follows: we use a camera to produce a single photo of a measured system --- as long as one can decompose such a system into two parts, one is performing a {\em simultaneous measurement} of those two parts. With respect to this definition, it is immaterial whether these two parts are two independent pendulums, or maybe two pieces of a single pendulum, or otherwise. By taking a single photo of the whole system, any two of its subsystems are said to have been photographed ``simultaneously''.

Another question one can ask regarding our experimental setup is the following. In our protocol, after taking a single photo of two pendulums we cut it into two pieces, one for each pendulum. Alternatively, we could have discussed a different protocol, in which we would simply take two photos, one for each pendulum, instead of taking one photo and cutting it in two. As it turns out, if one collects the data using this alternative protocol, one will fail to detect the extraordinary scatter plot (from Figure \ref{figure3}) when analysing the data. Taking a single photo and cutting it in two parts is thus a crucial feature of our protocol, one that enables us to detect the relevant signal. That is why we have deliberately chosen to introduce the protocol as was done in the previous Subsection.

A further issue that needs to be discussed is the question whether it is possible to obtain the desired signal in the case when the two pendulums are mutually interacting in some way. In order to address this question, let us consider an example, a situation with two electrically charged pendulums, so that they interact electromagnetically while swinging. Repeating the procedure of taking photos and analysing, like before, the new datasets $A$ and $B$, one obtains the results which differ from the ones above only in the details of the shape of the curve $\cC$. This is because the interaction between the two pendulums will influence their relative positions and motion, but the extraordinary permutation similar to the one in Figure \ref{figure3} will still exist, and the curve obtained from it will qualitatively remain the same, namely a $1$-dimensional curve. This is a generic result, independent of the type and details of the interaction between pendulums.

Next, in our case of the two pendulums the curve $\cC$ is almost certainly self-intersecting, preventing it from being a manifold in the proper sense. Second, if we denote the length of the pendulums $A$ and $B$ as $l_A$ and $l_B$ respectively, it may happen that their square-roots are incommensurate,
\begin{equation} \label{IncommensuratePendulums}
\sqrt{\frac{l_A}{l_B}} \notin \racionalni\,.
\end{equation}
Note that if we recall Newtonian mechanics, \eqref{IncommensuratePendulums} in fact means that the periods of the two pendulums are incommensurate. Of course, in order to be consistent, we have to assume that we have no knowledge about either Newtonian mechanics nor about the concept of a ``period'', so we refrain from this interpretation. Nevertheless, we are still allowed to measure the lengths $l_A$ and $l_B$ using a ruler, calculate their square roots, and discuss the validity of \eqref{IncommensuratePendulums} in a given experiment. And if it happens that \eqref{IncommensuratePendulums} holds, in the limit $N\to\infty$ the curve $\cC$ becomes a space-filling curve (passing through almost every point in the rectangle $S_A\times S_B$), invalidating the statement that it is $1$-dimensional, in the sense of the analysis discussed in Appendix \ref{AppA}. In order to work around these two issues, it is useful to enlarge our physical system yet again.

\subsection{Systems with three and more observables}

\begin{figure}[ht]
\begin{center}
\includegraphics[scale=0.5]{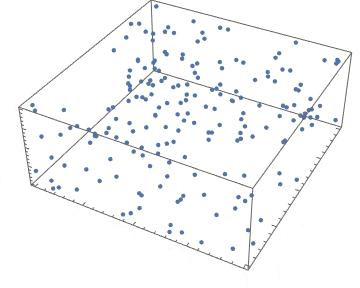}
\end{center}
\caption{\label{figure4}An ordinary scatter plot for a triple pendulum system.}
\end{figure}

\begin{figure}[ht]
\begin{center}
\includegraphics[scale=0.5]{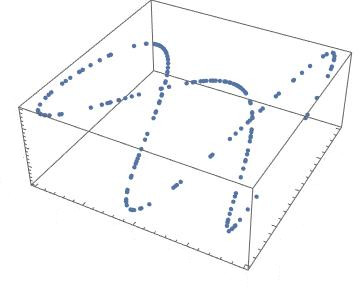}
\end{center}
\caption{\label{figure5}An extraordinary scatter plot for a triple pendulum system.}
\end{figure}

In addition to pendulums $A$ and $B$, let us introduce yet another pendulum, denoted $C$, and repeat the whole analysis, taking photos of all three pendulums side-by-side. This time we end up with the dataset
\begin{equation} \label{ThreePendulumDataset}
\begin{array}{lcl}
  A & = & \{ a_1, a_2, \dots, a_N \}\,, \\
  B & = & \{ b_1, b_2, \dots, b_N \}\,, \vphantom{\ds\int} \\
  C & = & \{ c_1, c_2, \dots, c_N \}\,. \\
\end{array}
\end{equation}
After sufficiently many measurements, we can establish the individual configuration spaces for each pendulum, as before,
\begin{equation}
\begin{array}{lclcl}
  S_A & \equiv & [a_{min},a_{max}] & \subset & \realni\,, \\
  S_B & \equiv & [b_{min},b_{max}] & \subset & \realni\,, \vphantom{\ds\int} \\
  S_C & \equiv & [c_{min},c_{max}] & \subset & \realni\,.
\end{array}
\end{equation}
We can now expect that the joint configuration space will be a measure-zero subset of the $3$-dimensional box $S_A\times S_B \times S_C$. To see this, introduce two arbitrary permutations of $N$ elements, $\pi$ and $\rho$, and construct the following $3\times N$ matrix,
\begin{equation}
\left(
\begin{array}{cccc}
a_1 & a_2 & \dots & a_N \\
b_{\pi(1)} & b_{\pi(2)} & \dots & b_{\pi(N)} \\
c_{\rho(1)} & c_{\rho(2)} & \dots & c_{\rho(N)} \\
\end{array}
\right)\,,
\end{equation}
which describes an arbitrary ordering of our dataset into triplets $(a_k,b_{\pi(k)},c_{\rho(k)})$, $k=1,\dots,N$. We use these triplets as data points in the $3$-dimensional scatter plot whose domain is the box $S_A\times S_B \times S_C$. One can construct $(N!)^2$ such scatter plots, one for each choice of the permutations $\pi$ and $\rho$. A typical plot will have data points randomly distributed and filling up the whole box, as in Figure \ref{figure4}.
However, either by using visual inspection over $(N!)^2$ such diagrams, or by using the data-analysis approach from Appendix \ref{AppA}, one can again establish the existence of the ``special'' diagram, see Figure \ref{figure5},
corresponding to the ``special'' permutations $\tilde{\pi}$ and $\tilde{\rho}$.

There are now several new features to be discussed. To begin with, as an experimental result of our thought-experiment, we can establish that our dataset $(A,B,C)$ together with the permutations $\tilde{\pi}$ and $\tilde{\rho}$, satisfies all three previously discussed properties of \textit{existence}, \textit{self-reinforcement} and \textit{dimensionality}. Therefore, the dataset again gives rise to a $1$-dimensional curve $\cC$, as an \textit{an experimental result}. In principle, the addition of the third pendulum could have rendered the data-points distributed along some $2$-dimensional surface in the box rather than a $1$-dimensional curve, while still being compatible with the $1$-dimensional curve if one only looks at the projection defined as the $(A,B)$ subset of the data. But such a scenario did not happen, and instead the data points are still aligned along a $1$-dimensional curve $\cC$. Therefore, it is a genuine experimental signal that the joint configuration space has $1$-dimensional structure. Moreover, this conclusion survives if one enlarges the physical system even further, by adding additional pendulums, or any other (even aperiodic) mechanical systems. From the point of view of our thought experiment, it is a completely general result.

Second, the $1$-dimensional nature of the curve $\cC$ is equivalent to the set of two correlation functions between three observables $a,b,c$,
\begin{equation}
F_1(a,b,c) = 0\,, \qquad F_2(a,b,c) = 0\,.
\end{equation}
In general, if we have a very big mechanical system of $K$ observables $a^{(1)},\dots ,a^{(K)}$, we will find in total $K-1$ correlation functions
\begin{equation} \label{GeneralMechCorrelationFunctions}
\begin{array}{lcl}
  F_1(a^{(1)},\dots, a^{(K)}) & = & 0\,, \\
 & \vdots & \\
  F_{K-1}(a^{(1)},\dots, a^{(K)}) & = & 0\,, \\
\end{array}
\end{equation}
whose set of solutions describes a $1$-dimensional curve $\cC$ in a $K$-dimensional configuration space.

Next, given the curve $\cC$, one can project it onto the $(a,b)$ plane, simply by ignoring the value of the observable $c$, so that one recovers a curve defined by the points $(a_k,b_{\tilde{\pi}(k)})$ in the space $S_A\times S_B$. Assuming that the datasets $A$ and $B$ are identical to those used in the previous, two-pendulum example, the ``special'' permutation $\tilde{\pi}$ that we have found in the previous subsection will be \textit{identical} to the one we found now. In other words, the presence or absence of the dataset $C$ in the statistical analysis of the Appendix \ref{AppA} will not change the permutation $\tilde{\pi}$, although a priori this is mathematically possible. The fact that this does not happen in our thought experiment is a consequence of the nontrivial nature of the dataset $(A,B,C)$.

Since the permutation $\tilde{\pi}$ is independent of the presence of the third dataset, we can use the third dataset to ``resolve'' the self-intersecting points of the projection of the curve $\cC$ to the space $S_A\times S_B$. Namely, the presence of the third observable establishes that we are actually observing a projection of two different points of the curve $\cC$ onto the same point in the $S_A\times S_B$ subspace. Specifically, if we have the following two points on the curve $\cC$,
\begin{equation}
\begin{array}{lcl}
  \cP_k & \equiv & (a_k,b_{\tilde{\pi}(k)},c_{\tilde{\rho}(k)})\,, \\
  \cP_m & \equiv & (a_m,b_{\tilde{\pi}(m)},c_{\tilde{\rho}(m)})\,, \vphantom{\ds\int} \\
\end{array}
\end{equation}
such that $a_k \approx a_m$, $b_{\tilde{\pi}(k)} \approx b_{\tilde{\pi}(m)}$ but $c_{\tilde{\rho}(k)} \neq c_{\tilde{\rho}(m)}$, we see that although the points $\cP_k$ and $\cP_m$ may belong to quite distant parts of the curve $\cC$, they will both project to the ``same point'' in the subspace $S_A\times S_B$, leading to the apparent intersection. This is again a general feature of mechanical systems --- if our curve $\cC$ happens to self-intersect anywhere, we can always enlarge the physical system by adding another observable which will ``distinguish'' between the two appearences of the apparent intersection point, resolving it into two distinct points.

Finally, let us just shortly note that the issue of space-filling curve $\cC$ can be eliminated by extending the physical system with an observable $x$ which has a noncompact domain $S_X$, and can be considered to be monotonically increasing, such as entropy or whatever else is at hand. This means that our big configuration space will be noncompact, and the curve $\cC$ will intersect every hypersurface $x=const$ precisely once, sidestepping any issue of space-filling curves.

\subsection{The time manifold}

The thought experiment described above provides an operational protocol to  establish the existence of a $1$-dimensional, non-self-intersecting, non-space-filling curve $\cC$ as a joint configuration space of a given number of observable mechanical degrees of freedom. The curve $\cC$ has all the hallmarks of a manifold (it is a nonempty set with a well-defined topology and dimension), and in order to properly promote it into a manifold, all we need is a set of coordinate charts, i.e., an \textit{atlas}. Assuming for simplicity that $\cC$ has a topology of an open line, it is enough to consider a single coordinate chart across the whole curve (as opposed to a circle which requires at least two charts), called {\em time chart}:
\begin{equation}
T: \cC \to \realni\,, \qquad (a,b,c,\dots) \mapsto t\,.
\end{equation}
Being a chart, i.e., a homeomorphism, this map is invertible, giving rise to \textit{parametric equations of motion} for the observables $a,b,c,\dots$,
\begin{equation}
T^{-1}: \realni\to \cC\,, \qquad t \mapsto (a(t),b(t),c(t),\dots)\,.
\end{equation}
This map defines $a(t)$, $b(t)$, etc., as functions of the \textit{time coordinate} $t\in\realni$, and they satisfy the correlation functions \eqref{GeneralMechCorrelationFunctions}. Of course, one can introduce a different chart $T': \cC \to \realni$, such that the composition $\phi \equiv T' \mycirc T^{-1}: \realni\to\realni$ is a homeomorphism of the real line. One can assume that $\phi$ is a smooth function, in which case it is a diffeomorphism. It defines the \textit{time reparametrization transformation}, as
\begin{equation}
t' = \phi(t)\,,
\end{equation}
which keeps the correlation functions \eqref{GeneralMechCorrelationFunctions} invariant. This becomes obvious if we use the composition notation $F_k(a(t),b(t),\dots) \equiv ( F_k \mycirc T^{-1} )(t)$, so that
\begin{equation}
F_k \mycirc T^{-1}  = 0
\end{equation}
obviously implies
\begin{equation}
 F_k \mycirc T^{-1} \mycirc \phi^{-1} = 0\,,
\end{equation}
for all $ k=1,2,\dots, K$, where $\phi^{-1}$ is a coordinate transformation from $t'$ to $t$. Since the curve $\cC$ is in fact the joint configuration space of $K$ observables, and it is defined by the equations $F_k = 0$, the invariance of the correlation functions establishes the invariance of the joint configuration space $\cC$ under the action of the $1$-dimensional group of diffeomorphisms $\Diff(\realni)$. This feature of $\cC$ is called the \textit{time reparametrization invariance}, and the group $\Diff(\realni)$ is a subgroup of the larger group of {\em spacetime diffeomorphisms}, $\Diff(\realni^4)$, as we shall see in the next section.

Let us finish this section with a remark that the existence of the curve $\cC$ in our dataset means that it is impossible to ``erase'' the information about coincident measurements of the observables. Recall that we were ``cutting the photos'' to display only individual pendulums, in order to erase the information about the pairing of experiment outcomes. The existence of the curve $\cC$ fully recovers that information from our dataset, which means that this information \textit{cannot be erased}. In other words, the information about ``conditional measurements'' --- measuring the position of the pendulum $A$ \textit{under the condition that} pendulum $B$ has some given position --- is itself observable, and is encoded in the correlations present in the dataset.

\section{\label{SecFieldTheory}Field theory and spacetime}

The generalization of the analysis given in the previous section, from a time manifold to a spacetime manifold, is completely straightforward. The only difference is that we need to use fields instead of mechanical systems in our thought experiment.

For the sake of concreteness, let us imagine that we have a fluid, flowing through some big container (a pipe or a river bed). Suppose we have an instrument, called ``a probe'', which we can insert into the fluid to measure its various properties. The probe has compact spherical shape and is small enough not to perturb the properties and flow of the fluid, while it samples (i.e., performs a coincident measurement of) various observables. When immersed into the fluid and activated, the probe provides the following set of numbers:
\begin{itemize}
\item mass-density $\rho^{m}$ of the fluid,
\item pressure $p$ of the fluid,
\item temperature $T$ of the fluid,
\item charge-density $\rho^{e}$ of the fluid,
\item magnitude $E \equiv \|\vec{E} \|$ of the electric field inside the fluid,
\item magnitude $B \equiv \|\vec{B} \|$ of the magnetic field inside the fluid, and
\item the angle $\theta$ between $\vec{E}$ and $\vec{B}$, specified as $\theta \equiv \vec{E}\cdot \vec{B} / EB$.
\end{itemize}
In total, the probe measures the $7$ observables
\begin{equation}
\rho^m, \quad p,\quad T,\quad \rho^e,\quad E,\quad B, \quad \theta \,.
\end{equation}
We then perform $N$ measurements, by using $N$ probes. Each probe randomly activates once, performs the measurements, and transmits the measured values wirelessly to our computer. In order to erase any information about ``where and when'' the measurements took place, the computer does not keep track of anything but the {\em unordered sets} of $N$ measurements for each observable,
\begin{equation}  \label{FTdataset}
\begin{array}{lcl}
\rho^m & = & \{ \rho^m_1,\dots,\rho^m_N \} \,, \\
p & = & \{ p_1,\dots,p_N \} \,, \\
 & \vdots & \\
\theta & = & \{ \theta_1,\dots,\theta_N \} \,. \\
\end{array}
\end{equation}
Each of the $7$ observables belongs to its domain, which is either a compact or a noncompact subset of $\realni$. Denote them in turn as
\begin{equation}
S_{\rho^m}\,, \qquad S_p\,, \qquad \dots\,, \qquad S_\theta\,,
\end{equation}
and introduce the total (kinematic) configuration space as the Cartesian product of all of these,
\begin{equation}
S_{kin} \equiv S_{\rho^m}\times S_p \times \dots\times S_\theta\,.
\end{equation}
Next, since we work with $7$ observables, introduce $6$ arbitrary permutations $\pi_1,\dots,\pi_6$ of a set of $N$ elements, and construct a $7\times N$ matrix as
\begin{equation}
\left(
\begin{array}{cccc}
\rho^m_1 & \rho^m_2 & \dots & \rho^m_N \\
p_{\pi_1(1)} & p_{\pi_1(2)} & \dots & p_{\pi_1(N)} \\
 & & \ddots & \\
\theta_{\pi_6(1)} & \theta_{\pi_6(2)} & \dots & \theta_{\pi_6(N)} \\
\end{array}
\right)\,.
\end{equation}
Each column in this matrix represents one data point in a $7$-dimensional scatter plot, which obviously cannot be drawn on paper but is well-defined nevertheless. Since the permutations $\pi_1,\dots,\pi_6$ can be chosen completely arbitrary, there are in total $(N!)^6$ such matrices, each with its own $7$-dimensional scatter plot. Again for obvious reasons, visual inspection of all these plots is not possible, but the statistical analysis from Appendix \ref{AppA} should work just fine, provided enough computational power. As a result of the thought experiment, the statistical analysis will provide us with one ``special'' choice of permutations $\tilde{\pi}_1,\dots,\tilde{\pi}_6$, for which all data points align themselves nicely along one hypersurface, denoted $\cM$, in the big $7$-dimensional configuration space $S_{kin}$. In the limit $N\to\infty$, it satisfies the properties of being a strict measure-zero subset of $S_{kin}$:
\begin{equation}
\cM \subset S_{kin}\,, \qquad \frac{\meas(\cM)}{\meas(S_{kin})} = 0\,.
\end{equation}

The hypersurface $\cM$ represents the joint configuration space for our physical system, and exhibits the usual properties of \textit{existence}, \textit{self-reinforcement} and \textit{dimensionality}, all established as an experimental result of our thought experiment. The statistically obtained dimension of $\cM$, without any a priori reason whatsoever, turns out to be
\begin{equation} \label{eq:SpacetimeDimResult}
\dim \cM = 4\,.
\end{equation}
Note that this conclusion about the dimensionality, as well as any conclusion based on experimental data, is ultimately contingent on the precision of the equipment used to obtain that data, and may change with the advance of technology and science. See Subsection \ref{SubSec:ExtraDim} for a detailed discussion of this issue. See also Subsection \ref{SubSec:emergence} for a detailed discussion of various theoretical proposals to calculate the dimension of spacetime from first principles, such as spinfoam models, string theory, causal set theory, and others.

Also, as before, if the hypersurface $\cM$ happens to self-intersect or fills up the whole space $S_{kin}$, we should simply add additional convenient observables to our dataset, and convince ourselves that the above inconvenient properties dissapear. Also as before, adding additional observables fails to change the overall dimension of $\cM$, which persistently keeps being equal to $4$, regardless of the number of sampled observables.

If we extend the number of our observables from $7$ to $K$, denoted $a^{(1)},\dots,a^{(K)}$, we find a total of $K-4$ correlation functions
\begin{equation} \label{GeneralFTCorrelationFunctions}
\begin{array}{lcl}
  F_1(a^{(1)},\dots, a^{(K)}) & = & 0\,, \\
  & \vdots & \\
  F_{K-4}(a^{(1)},\dots, a^{(K)}) & = & 0\,, \\
\end{array}
\end{equation}
whose set of solutions describes a $4$-dimensional hypersurface $\cM$ in a $K$-dimensional configuration space.

The only thing left to do at this point is to introduce a set of coordinate charts, i.e., an atlas,
\begin{equation}
\begin{array}{lccc} 
  f: & \cM & \to & \realni^4\,,  \\
  & (a^{(1)},\dots,a^{(K)}) & \mapsto & x \equiv (t,x,y,z)\,, \vphantom{\prod^A\limits} \\
\end{array}
\end{equation}
and its inverse
\begin{equation}
\begin{array}{lccc}
  f^{-1}: & \realni^4 & \to & \cM \,, \\
  & x & \mapsto & (a^{(1)}(x),\dots,a^{(K)}(x)) \,, \vphantom{\prod^A\limits} \\
\end{array}
\end{equation}
which establish the parametric functions $a^{(k)}(x)$ for the observables, which in turn satisfy all correlation functions \eqref{GeneralFTCorrelationFunctions}. The hypersurface $\cM$, now established as a proper $4$-dimensional manifold, is of course called \textit{spacetime}, while the parametric functions $a^{(k)}(x)$ are called \textit{fields living on spacetime}.

One can also introduce a different chart,
\begin{equation}
\begin{array}{lccc}
  f': & \cM & \to & \realni^4\,, \\
  & (a^{(1)},\dots,a^{(K)}) \!\! & \mapsto & \!\! x' \equiv (t',x',y',z')\,, \vphantom{\prod^A\limits} \\
\end{array}
\end{equation}
such that the composition $\phi = f'\mycirc f^{-1} : \realni^4 \to \realni^4$ is a homeomorphism in $\realni^4$. If $\phi$ is smooth, it is a diffeomorphism. It defines a \textit{coordinate transformation}
\begin{equation}
x' = \phi(x)\,,
\end{equation}
which keeps the correlation functions \eqref{GeneralFTCorrelationFunctions} invariant, similar to the $1$-dimensional case from Section \ref{SecMechanics}. Since $\cM$ is equivalent to the solution of the system \eqref{GeneralFTCorrelationFunctions}, and since it is the joint configuration space for the observables, it is invariant under the action of the group of $4$-dimensional diffeomorphisms, $\Diff(\realni^4)$. This feature of $\cM$ is called (passive) \textit{diffeomorphism invariance of spacetime}, or \textit{general coordinate invariance}.

Since $\cM$ is the joint configuration space for the observables, choosing a particular data point $k$ in spacetime (one of the measured points on $\cM$) gives rise to a $K$-tuple of particular values of the observables. In our example, those are
\begin{equation}
( \rho^m_k, p_{\tilde{\pi}_1(k)}, \dots, \theta_{\tilde{\pi}_6(k)} )\,,
\end{equation}
and they reconstruct the information about coincident measurements coming from each particular probe --- the information we had tried to erase by ignoring any particular order of values in the dataset \eqref{FTdataset}. As in the case of the $1$-dimensional time manifold, this information is present in the correlations of the dataset itself, and \textit{cannot be erased}.

As a final point, note that the above operational reconstruction of spacetime is actually one precise implementation of the relationalism paradigm, defining the spacetime manifold using nothing but fields that supposedly live on it. The idea for a mental image ``fields do not live on top of spacetime, but on top of each other'' (pointed out for example by Rovelli in \cite{Rovelli2004}) can be explicitly realized if we were to take $4$ parametric functions $a^{(1)}(x),\dots,a^{(4)}(x)$, conveniently chosen so that they uniquely specify a spacetime point $x$ in some given chart, i.e., such that one can solve those $4$ parametric functions for the coordinates as functions of the observables,
\begin{equation}
x^{\mu} (a^{(1)},\dots,a^{(4)})\,, \qquad \mu=0,\dots,3\,,
\end{equation}
and then use them to eliminate $x$ from the remaining $K-4$ observables $a^{(5)},\dots,a^{(K)}$. In this way one arrives precisely at the $K-4$ correlation functions \eqref{GeneralFTCorrelationFunctions}, which actually define the manifold $\cM$ using nothing but the information about fields. These correlation functions are the precise technical implementation of the statement that ``fields live on fields'', in the sense that we can only observe coincidences among fields, as opposed to the the values of fields ``at a given spacetime point''.

Nevertheless, the fact that we always have precisely $K-4$ correlation functions (that is, for every choice of $K$) tells us that the set of solutions of those $K-4$ correlation functions, namely the spacetime manifold $\cM$, is itself operationally observable, independent of the choice and the amount of the fields one uses to describe it, and despite the diffeomorphism symmetry of those fields. The invariant properties of $\cM$ such as dimension and topology are present as correlations in the experimental data, and there is no theoretical account of why they have the values that we observe in the experiment. The $4$-dimensionality and simply connected topology are {\em brute experimental facts}, and are {\em independent} of the choice, properties and even the very number of the fields we use to measure them (given current science and technology, as mentioned below equation \eqref{eq:SpacetimeDimResult} and discussed in detail in Subsection \ref{SubSec:ExtraDim}). Of course, one could attempt to construct a theoretical model which would be able to deduce these correlations among observables from some simpler set of principles. This would implement the concept of the {\em emergence of spacetime}. So far, however, no such model has ever been successfully constructed.


\section{\label{SecConclusions}Conclusions and discussion}

In this work, we have argued that one can give an operational, model independent experimental protocol whose outcome would be the determination of the dimension and the topology of the time manifold (Section \ref{SecMechanics}) and the spacetime manifold (Section \ref{SecFieldTheory}). After introducing the thought experiment, we have studied the relevant properties (see Conjecture 2 from Section \ref{SecMechanics}) of the experimental data which give rise to a signal that reflects the observability of the underlying manifold, over which the fields are defined.

In what follows, we will discuss various aspects of the thought experiment that were not discussed in detail in previous sections, but nevertheless deserve to be mentioned and commented on.

\subsection{Distinguishing space from spacetime}

In Section \ref{SecMechanics} we have introduced the time manifold by looking at a mechanical system, while in Section \ref{SecFieldTheory} we have introduced the spacetime manifold, looking at fields instead. However, one may ask a natural question about the notion of a space manifold, and its differences from spacetime. To that end, let us discuss another illustrative example of the application of our gedanken-eksperiment.

Consider a room with a lamp and some furniture. The experimental apparatus used to perform measurements over this system consists of two cameras separated several centimeters apart. Each camera performs a measurement that provides the following data:
\begin{itemize}
\item the polar and azimuthal angles $\theta$ and $\varphi$ of the incident light ray,
\item the frequency $\nu$ and intensity $I$ of the ray.
\end{itemize}
Since we have two cameras, in total there are eight observables per measurement. As always, we collect the measured data ignoring any order, and perform the analysis of the protocol, to reach the conclusion that there are five correlation functions between observables, which means that all measurements can be arranged on a 3-dimensional manifold. Intuitively, this camera setup corresponds to the stereoscopic eyesight, that gives one a perception of depth and thus the notion of a 3-dimensional space of the room. One should note that the notion of time is absent from this description, since the scene of the room is static. Hence the resulting 3-dimensional manifold deserves the name {\em space manifold}.

Nevertheless, one can also consider a room with a lamp and furniture, and additionally a cat moving around inside. If we now perform the same type of measurements of the system with identical cameras as before, we will find only four correlation functions between eight observables, which means that all measurements can be arranged on a 4-dimensional manifold, rather than a 3-dimensional one. It is obvious that the motion of the cat renders the dataset fundamentally more complicated, with less amount of correlation. In other words, the observed 4-dimensional manifold corresponds to spacetime, since in this case the scene of the room is not static anymore.

This example illustrates the dependence of the outcome of the experiment on the properties of the system being observed. It may happen that a system has a high level of symmetry (in the example above, the time-translation invariance), which introduces additional correlations into the dataset and lowers the dimension of the resulting manifold. This is the main way to distinguish space from spacetime --- space is in fact spacetime with a certain global symmetry, which renders time unobservable. Similarly, the time manifold from the pendulum example in Section \ref{SecMechanics} is also spacetime with a global space-translation symmetry (the pendulum swings the same way regardless of its spatial position), which renders space unobservable.

\subsection{Quantum-mechanical treatment}

It is not completely obvious what would change regarding the outcomes of our thought experiment, if one were to take into account the effects of non-commuting observables, i.e., quantum effects. On one hand, the above analysis makes use of only mutually commuting observables (positions of several pendulums in mechanics, or values of different fields in field theory all mutually commute). This may suggest that our results should not be disturbed by the fact that there exist other observables, which fail to commute with the ones used in the experiment. In principle, one can extend the set of sampled observables up to the so-called {\em complete set of compatible observables}, without changing anything in the above analysis.

On the other hand, a very precise measurement of a coordinate of a given pendulum may uncontrollably perturb its momentum, so much that the pendulum fails to swing in the usual way, which would introduce a form of an intrinsic noise in the expermental data. In that case, subsequent measurements of the position of the pendulum may fail to be well correlated to the measurements of other observed pendulums, potentially hindering the predictions for the dimension and topology of the proper configuration space. Indeed, whenever the Hamiltonian does not commute with the position operator, a precise position measurement inherently disturbs the system’s subsequent dynamics, resulting in potentially unbounded disturbances. This should be taken into account in any generalisation of our gedanken-experiment to the quantum realm.

Additionally, care should be taken to distinguish the single-shot measurement of a single instance of a pendulum (which gives a random result, per QM), from the statistical measurement of an ensemble of identically prepared pendulums (which is probabilistically determined by QM). The uncertainty relations hold for the latter, while in our study we are interested in the former.

The proper quantum mechanical treatment is out of the scope of this paper, and we postpone it for future work.

\subsection{\label{SubSec:ExtraDim}Extra dimensions of spacetime}

Throughout Section \ref{SecFieldTheory}, it was claimed that the correlations one ought to find in real experimental data will ultimately support the conclusion $D \equiv \dim \cM = 4$. This claim is supported by everyday experience and virtually all experiments ever performed in the history of physics so far. These experiments roughly cover the scales from $10^{-20}\,{\rm m}$ (the scale of the current LHC and LIGO experiments), up to $10^{26}\,{\rm m}$ (the scale of the observable Universe). There is a further range of scales, from $10^{-20}\,{\rm m}$ down to $10^{-35}\,{\rm m}$ (the Planck scale), and maybe even beyond that. We currently have no experimental data from this range to either support the result $D=4$ or disprove it. Eventual access to this data could in principle change the result for $D$. For example, in the context of string theory \cite{PolchinskiBook1,PolchinskiBook2}, one imagines spacetime to have six additional spacelike dimensions, wrapped up into a small Calabi-Yau manifold. If we were to measure various fields at the scale smaller than the size of that Calabi-Yau manifold, our analysis of the data would yield $D=10$, or $4$ plus whatever number of compactified small extra dimensions exist at that scale.

In this sense, the dimension of the spacetime manifold may be a scale-dependent quantity, like running coupling constants in QFT, having different ``effective'' values at different scales. So far we have no data that would indicate anything other than $D=4$, but in principle this may change. Related to this, one may ask if our analysis supports the ``running'' of $D$ to values smaller than $4$ at smaller scales. Values larger than $4$ are obviously possible, but smaller values are also possible in principle. Namely, with sufficient resolution, one may notice that what looks like a $4$-manifold is (for example) in fact a $1$-dimensional curve densely packed up in a space-filling fashion, like some finite iteration of the Peano curve. In this sense, a fundamentally $1$-dimensional manifold can look at large scales as a $4$-dimensional one. Something along these lines apparently happens in the context of Causal Dynamical Triangulations~(CDT) scenario~\cite{CDTreview1,CDTreview2}, giving rise to $D=2$ near the Planck scale. That said, note that the CDT results discuss a different concept of a spacetime dimension, namely an effective dimension visible to a random walker. This is in general not equivalent to the notion of topological dimension that we discuss here. Namely, in the context of chaos theory, one typically introduces the notion of a so-called fractal or Hausdorff dimension, which essentially captures the effective dimension visible to a random walker exploring the fractal structure. In general, this notion is different from the ordinary, topological definition of dimension, that we use in this work (see equation \eqref{StatisticalDimensionDef} from Appendix \ref{AppA}). Nevertheless, in cases where the geometric structure under investigation has appropriately 'nice' features, the notions of the Hausdorff and topological dimensions coincide, and give an integer value that corresponds to our intuitive notion of dimension (see \cite{Falconer} for further details). Another context in which one often discusses different effective values of dimension at different scales is Causal Set Theory (CST) \cite{CSTreviewPaper}. Namely, in this scenario, the fundamental structure describing spacetime is a discrete set of points, which ultimately has topological dimension zero, but in a semiclassical limit gives rise to general relativity and the traditional notion of a $4$-dimensional smooth manifold as an effective description on large scales. At the conceptual level, the running of the dimension from $4$ to $0$ could also be observed using our technique, provided that one can perform measurements at the scale where spacetime discreteness is dominant. Unfortunately, we have no technological capabilities to do this in practice. Nevertheless, the relationship between different dimensions of the manifold at different scales is also well supported theoretically in the CST context, see for example \cite{Butterfield}.

The result $D=4$ is also contingent on the choice of observables measured and used to conclude that $D=4$. In principle, if we were to extend our set of observables, we could find additional \textit{large} dimensions. For example, so far we can detect the presence of dark matter only through its gravitational interaction with regular matter. In addition to that, one could imagine that there are also direct (contact) interactions between dark and regular matter, but that they are obscure enough not to be easily visible (like neutrinos, which interact with other matter only through short-range weak interactions and even weaker gravity). Nevertheless, if we somehow manage to measure these additional observables dependent on dark matter, they may change our statistics even in the IR regime, i.e., at large scales. It may thus turn out that our analysis yields $D=7$ or $D=12$ or any number larger than $4$. An example of this scenarios are braneworld cosmologies similar to Randall–Sundrum model \cite{RandallSundrum1,RandallSundrum2} and similar, where all ``standard'' observables we can measure happen to be nonzero only on a $4$-dimensional submanifold of some target manifold of larger dimension, while the ``dark matter'' observables would be nonzero even in the bulk of the target manifold, ultimately giving rise to $D>4$. So far we have not found any such observables, but this may also change in the future.

\subsection{Uniqueness objections}

Looking at the analysis of the dataset of our thought experiment, one can raise an objection that the analysis of the correlations between observables may fail to give a unique result. Namely, in generic circumstances, any finite dataset constructed by $N$ measurements of $K$ observables can be permuted into $(N!)^{K-1}$ possible arrangements, and for each of them one can evaluate the suitable critical parameter (see Appendix \ref{AppA}) that singles out the one ``special'' permutation. Since the set of permutations is finite, we end up with a finite collections of the values of the critical parameter, and any such collection contains a minimal element. This minimal parameter corresponds to a ``special'' permutation, since it features the strongest correlation in the dataset, ultimately describing a spacetime manifold. Given this setup, it may turn out that this minimal parameter is not unique, in the sense that several different permutations of the dataset all have this same minimal value of the parameter. In such a case, there are in principle several different possible arrangements of data featuring equally strong correlations, leading to several different possible candidate manifolds.

On one hand, it is not feasible to study this question numerically in practice, since the set of $(N!)^{K-1}$ of all possible permutations is incredibly huge, while permutations featuring the minimal parameter are likely to be a scarce subset of these. This limits us to theoretical arguments that generic datasets feature minimal critical parameter for only one permutation. The main argument provided so far is based on the \textit{self-reinforcement} property, which is unlikely to hold for more than one permutation and its restrictions to arbitrary subsets of data.

On the other hand, it is easy to construct explicit examples which feature more than one ``special'' permutation, each corresponding to the same minimal value of the critical parameter. This is expected to happen if a physical system being measured features global symmetries. For example, pendulums are invariant with respect to a global left-right symmetry, in the sense that one can find two ``special'' permutations, which are mirror images of each other. While these formally give rise to two different time manifolds, these manifolds are represented by two isomorphic datasets, and therefore in fact represent the one and the same time manifold, despite the nonuniqueness. In this sense, nonuniqueness that stems from the existence of global symmetries is benign, and does not invalidate the analysis and conclusions of the thought experiment.

Also, one of the most often discussed examples is the ``diagonal'' permutation of the dataset, where the values for each observable are sorted in an ascending (or descending) fashion. This leads to a dataset where points are roughly aligned along a $1$-dimensional curve connecting two opposite corners of the $K$-dimensional kinematic configuration space. Being $1$-dimensional, this permutation is likely to have a very small value of the critical parameter, possibly the minimal one. Nevertheless, it is not hard to demonstrate that this permutation fails to satisfy the self-reinforcement property, since its restrictions to data subsets will fail to coincide with the full permutation. In other words, as one adds more data to the dataset, the resulting $1$-dimensional curve ``wiggles'', changing its shape for every extension of the dataset. Thus such a permutation has to be excluded from consideration, regardless of the very small value of the critical parameter. We conjecture that the self-reinforcement property will be violated for all permutations featuring small critical parameter, except for one --- which can thus be uniquely recongized as ``special''. However, any potential proof of this conjecture is not available at this time, and requires further study.

\subsection{\label{SubSec:emergence}Emergence of spacetime}

As we noted at the end of Section \ref{SecFieldTheory}, in principle one can imagine a theoretical model which does not feature anything like a $4$-manifold in its postulated structure. Further, such a model may give us predictions for the values of all possible observables, and we can use it to generate a dataset, study it using our methods, and end up with correlations among observables which give rise to a $4$-manifold. If such a scenario happens, one says that spacetime {\em emerges} from the theoretical model, and that the model predicts the values of spacetime dimension and topology.

Despite many hopeful attempts (usually in the context of quantum gravity), there are no particular theoretical models that have managed to achieve this, even with a wrong prediction for the dimension and topology. The reason for this are two important criteria that such a model must satisfy in order to make a legitimate prediction:
\begin{itemize}
\item[(1)] The information about the $4$-manifold must not be present in the axiomatic structure of the model. Namely, if we include the $4$-manifold structure as an input in the construction of the theory, it should come as no surprise that one can recover that information from the model later on, in various different forms. However, this can be tricky to test, since the information about a $4$-manifold can be encoded in a non-obvious, cryptic fashion, and it may be hard to prove that some of the axioms of the model are indeed equivalent to the assumption that there exists a $4$-manifold in the theory.
\item[(2)] One must demonstrate that the actual dimension of the would-be manifold is explicitly computable from the model. For example, one can try to evaluate a bunch of observables using the model, and then apply our analysis onto that data, in order to obtain $D=4$. In this sense, the dimension of spacetime would be an explicit consequence of the dynamics of the observables in the model. Alternatively, one may use some other way to calculate the spacetime dimension, but this again needs to be a consequence of dynamics of the observables, giving rise to appropriate correlations in the dataset that would ultimately lead to a $4$-manifold. However, it is not enough to handwavingly claim that the model ``in principle'' predicts these correlations, because they cannot be a generic feature of the model. The correlations must be \textit{explicitly calculated}, or otherwise rigorously proved to exist, and in addition it must be also rigorously proved that there cannot be any further correlations beyond these, since any further correlations would lower the dimension below $4$, falsifying the model.
\end{itemize}

For example, a typical spinfoam model of quantum gravity is constructed as a state sum over the values of the fields living on a $2$-complex (the spin foam), providing one with a way to calculate (expectation) values of observables \cite{EPRL,FK}. However, usually by construction, the $2$-complex which is used is dual to a triangulation of a $4$-manifold. Because of this property, the spinfoam model fails to satisfy the criterion (1) --- the information about a $4$-manifold is already integrated into the model as an axiom, rather than being a property of the dynamics of the observables. Moreover, if one manages to ``fix'' this by generalizing the $2$-complex structure somehow, so that it fails to be dual to a manifold (for example, by arranging that each dual cell has different topological dimension), there is criterion (2) --- one must use the model to explicitly demonstrate that the observables will always feature appropriate correlations (as a consequence of the dynamics of the model) so that they give rise to a $4$-manifold, at least in some large-distance limit. No such model has ever been proposed, and no such calculation has ever been performed.

Another example would be bosonic string theory, which is formulated by explicitly assuming the existence of a $D$-manifold, albeit with an unspecified value for $D$ \cite{PolchinskiBook1,PolchinskiBook2}. Then one uses the dynamics of the model to evaluate $1$-loop beta-functions, and from the requirement that these vanish, one obtains a consistency requirement $D=26$. The explicit assumption of a $D$-dimensional manifold is an input to the model, and thus already violates criterion (1). Criterion (2) comes very close to being satisfied, but unfortunately it is contingent on the particular choice of $\zeta$-function regularization, making use of the famous ``identity''
\begin{equation}
\zeta(-1) \equiv \sum_{n=1}^{\infty} n \equiv 1 + 2 + 3 + \dots = -\frac{1}{12}\,,
\end{equation}
obtained by analytic continuation of the Riemann $\zeta$-function, i.e., postulated as an axiom of the model. This ultimately renders the value of $D$ to be a part of the definition of the theory, completing the argument that criterion (1) is indeed violated.

A third example would be causal set theory \cite{CSTreviewPaper}. Here, the fundamental structure of spacetime is assumed to be a discrete set of points, without any assumption about $4$-dimensional manifold structure. In this sense, such models of quantum gravity indeed seem to satisfy our criterion (1). Moreover, when studying the semiclassical limit of such a model, one can go quite far in obtaining a classical theory of general relativity on a smooth manifold structure, see for example \cite{DowkerButterfield}. In this paper, there is even a discussion about the derivation of the spacetime dimension from the fundamental causal set structure, using various dimension estimators and discriminators. However, any explicit calculation involving those estimators that would uniquely provide the result $D=4$ is still missing. Thus, the so far discussed causal set models of quantum gravity, or otherwise any future proposals, are yet to satisfy our criterion (2).

One can naturally expect to find many other proposals for spacetime emergence {\em from various theoretical models} throughout the literature. However, in order to take any of these proposals seriously, one must first demonstrate that criteria (1) and (2) are fulfilled, which is highly nontrivial, and likely has not been done for any of the existing proposals.

Let us note that one might also consider another type of emergence, which stems {\em from raw experimental data}, as opposed to emergence {\em from a theoretical model}, which was discussed so far in this Subsection. In fact, the approach taken in this work represents an example of this second type of spacetime emergence. In that sense, the spacetime emergence from raw experimental data is precisely a synonym for the notion of the verification of the existence of spacetime, as used in the title. It is important to emphasise that the term ``emergence'' can thus have quite different meanings, depending on the context, and one should take care in its use.

\medskip

\acknowledgments

The authors wish to thank Jovan Janji\'c for help in implementing numerical algorithms in Mathematica, and to Igor Salom, \v Caslav Brukner, Marcus Huber, Reinhard Werner and Klaus Fredenhagen for fruitful discussions.

This work is funded by FCT/MECI through national funds and, when applicable, co-funded EU funds under UID/5008: Instituto de Telecomunica\c{c}\~oes. NP acknowledges the FCT Est\'{i}mulo ao Emprego Cient\'{i}fico grant no. CEECIND/04594/2017/CP1393/CT000, as well as the FCT project QuantumPrime reference PTDC/EEI-TEL/8017/2020.

MV was supported by the Ministry of Education, Science and Technological Development of the Republic of Serbia, and by the Science Fund of the Republic of Serbia, grant 7745968, ``Quantum Gravity from Higher Gauge Theory 2021'' --- QGHG-2021. The contents of this publication are the sole responsibility of the authors and can in no way be taken to reflect the views of the Science Fund of the Republic of Serbia.

\onecolumn
\appendix

\section{\label{AppA}Statistical analysis technique}

Suppose we are given a dataset of $N\to\infty$ points as $K$-tuples in some compact volume of a $K$-dimensional Euclidean space (the noncompact case can be reduced to the compact case by cutting it into countably infinitely many compact pieces, and studying each piece one by one). The volume of the box is known to be $V_K$. Our task is to verify whether, and to what extent, these $N$ points are ``nicely'' aligned on some $D$-dimensional hypersurface in the big $K$-dimensional space. Obviously, it is assumed that $D<K$, and the notion of ``nice'' alignment will be defined rigorously below.

\subsection{The critical parameter}

We proceed as follows. Pick any datapoint $x$, and construct around it a $K$-dimensional cube, such that the point $x$ is in the cube's center. Let the length of each edge of the cube be specified as
\begin{equation} \label{SizeOfTheCube}
\epsilon = \sqrt[K]{\frac{V_K}{N}}\,,
\end{equation}
i.e., such that the volume of the cube is
\begin{equation} \label{VolumeOfAsingleCube}
V_{cube} \equiv \epsilon^K = \frac{V_K}{N}\,.
\end{equation}
Suppose first that datapoints are scattered randomly throughout the box. If we construct the same cube around every datapoint, the total volume of all cubes can be represented as:
\begin{equation} \label{VtotalDef}
V_{total} = \sum_{n=1}^N \frac{(-1)^{n+1}}{n!} \sum_{i_1\neq\dots\neq i_n} V_{i_1\dots i_n}\,.
\end{equation}
Here, $V_{i_1\dots i_n}$ is the volume of the joint intersection of $n$ cubes labeled by indices $i_1,\dots,i_n$, and the second sum is evaluated over all nonrepeating values of these indices, which take values from the set $\{1,\dots,N\}$. Given that all individual cubes are of the same volume, the term for $n=1$ is
\begin{equation}
V_{n=1} = \sum_{i=1}^N V_i = NV_{cube} = V_K\,.
\end{equation}
Up to an overall minus sign, the remaining terms in \eqref{VtotalDef} are collected into a total {\em overlap volume}
\begin{equation}
V_{overlap} \equiv - \sum_{n=2}^N \frac{(-1)^{n+1}}{n!} \sum_{i_1\neq\dots\neq i_n} V_{i_1\dots i_n}\,,
\end{equation}
so that we have
\begin{equation} \label{VtotalDefDva}
V_{total} = V_K - V_{overlap} = \alpha V_K\,.
\end{equation}
The coefficient $\alpha \in [0,1]$ is another measure of this overlap:
\begin{equation} \label{AlphaDef}
\alpha = 1- \frac{V_{overlap}}{V_K} \,.
\end{equation}
In the main text, we call $\alpha$ the {\em critical parameter}.

Specifically, if the datapoints are very uniformly distributed throughout the box, their corresponding cubes will have very little overlap volume, $V_{overlap} \approx 0$, so that the volume covered by all cubes will be approximately equal to the volume of the box, $V_K$, and $\alpha \approx 1$. On the other hand, one can suppose that the datapoints are all aligned along some $D$-dimensional hypersurface $\cM$, whose total $D$-dimensional volume is $V_D$, while its $K$-dimensional volume is of course of measure zero. Now, if all $N$ datapoints are aligned on $\cM$, their cubes will overlap quite a lot, giving $\alpha \approx 0$ in the limit $N\to\infty$.

\begin{figure}[ht]
\begin{center}
\includegraphics[scale=0.3]{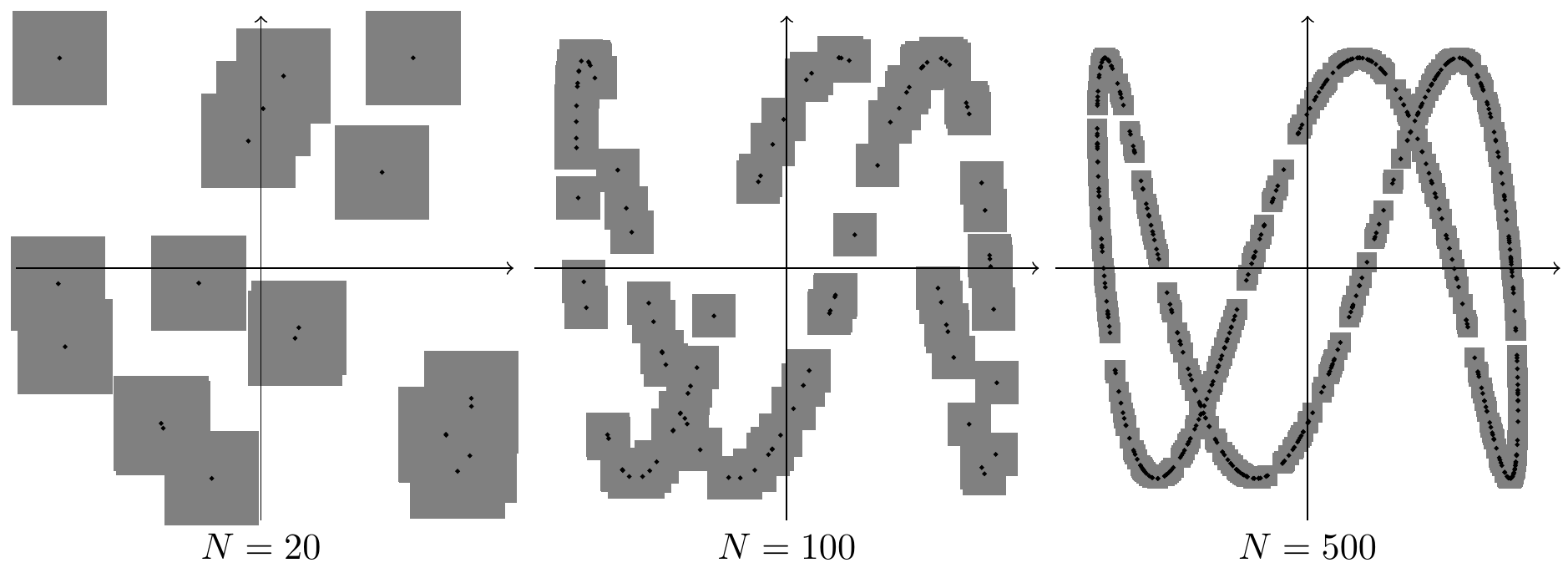}
\end{center}
\caption{\label{figure6}Successive approximations of the datasets, consisting of $N$ points, with a manifold-like structure.}
\end{figure}

\begin{figure}[ht]
\begin{center}
\includegraphics[scale=0.3]{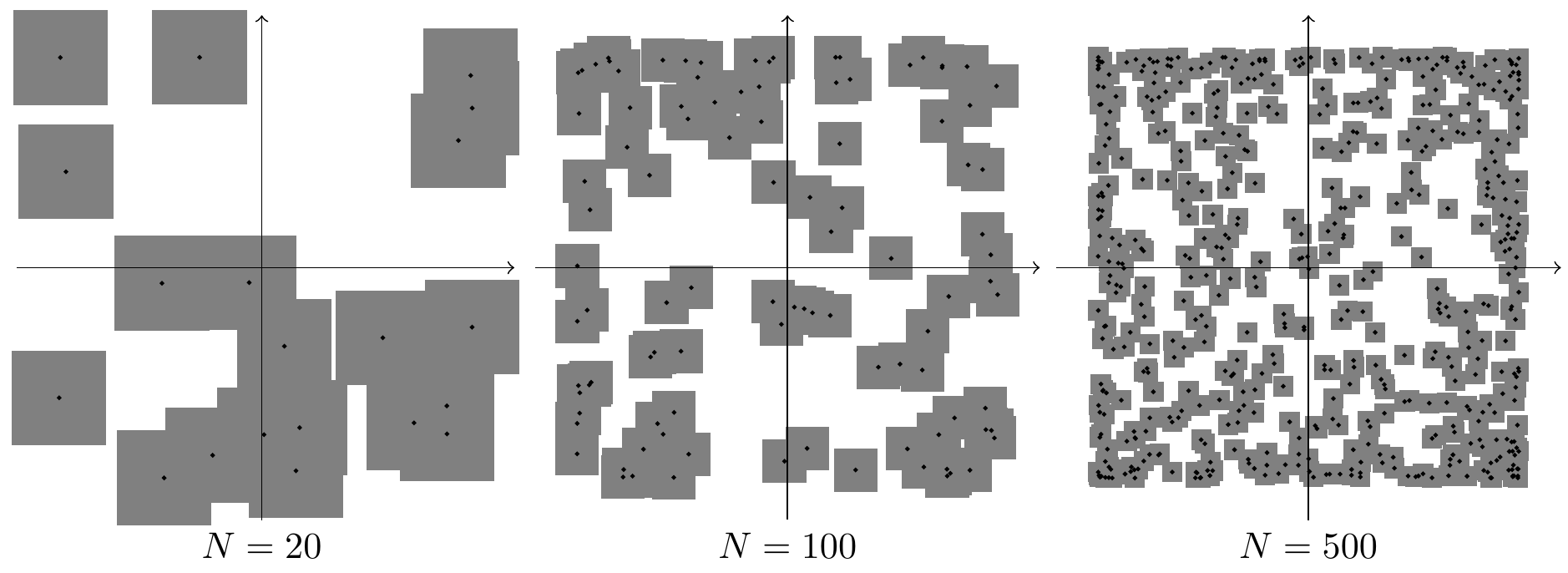}
\end{center}
\caption{\label{figure7}Successive approximations of the datasets, consisting of $N$ points, with a non-manifold-like structure.}
\end{figure}

One can estimate how close $\alpha$ is to zero in the following way. Construct a cube around every point lying on $\cM$. The intersection between $\cM$ and the cube will be again a $D$-dimensional hypersurface, while the cubes will ``protrude out'' of $\cM$ into $K-D$ orthogonal directions. We end up with a ``thick hypersurface'', having thickness $\epsilon^{K-D}$. Since by assumption all $N$ points are lying on $\cM$, one can approximate the value of $V_{total}$ of this thick hypersufrace as
\begin{equation}
V_{total} \approx \epsilon^{K-D} V_D = \left( \frac{V_K}{N} \right)^{\frac{K-D}{K}} V_D\,.
\end{equation}
This gives us the following estimate for $\alpha$:
\begin{equation} \label{GoodAsymptotics}
\alpha(N) \equiv \frac{V_{total}}{V_K} \approx \frac{V_D}{(\sqrt[K]{V_K})^D} N^{\frac{D}{K}-1} = const \cdot N^{\frac{D}{K}-1} \,.
\end{equation}
Since $D<K$, we see that in the limit $N\to\infty$ we have $\alpha\to 0$. Pictorially, the ``thickness'' $\epsilon^{K-D}$ of the hypersurface gradually shrinks as $N$ grows, due to \eqref{SizeOfTheCube}, so that in the limit $N\to\infty$ we have $\epsilon\to 0$, and our thick hypersurface becomes asymptotically infinitely thin. This is all illustrated for the case $K=2$ in the diagrams given in Figure \ref{figure6}.
On the other hand, if we take a random permutation of the second coordinates, the resulting diagrams look as the ones from Figure \ref{figure7}.
We can observe that, as $N$ grows, the area covered by the gray squares becomes smaller in the first case, while it remains approximately constant in the second (permuted) case. In the limit $N\to\infty$ the gray area drops to zero in the first case, while it remains constant in the second case.

\subsection{The asymptotic behavior}

Let us now formalize the procedure for the statistical analysis. Given the dataset of $N$ points and the volume $V_K$ of the big space, a computer can actually evaluate $V_{total}$ by explicit numerical evaluation of the overlap $V^{(n)}_{overlap}$ between the cube constructed around each point $n$ and all surrounding cubes. This means that $\alpha(N)$ can be explicitly evaluated, for each $N$. Provided that we have evaluated it for a number of different values of $N$, we can discuss its asymptotic structure in the limit $N \gg 1$ (which is formally equivalent to $N\to\infty$), as follows:
\begin{equation} \label{GeneralAsymptotics}
\alpha(N) = c_K + \sum_{n=0}^{K-1} c_n \frac{1}{(\sqrt[K]{N})^{K-n}} + \cO\left( \frac{1}{(\sqrt[K]{N}) ^{K+1} } \right) + \cR(N)\,.
\end{equation}
Here we have expanded $\alpha(N)$ into an asymptotic power series, from the order $1/\sqrt[K]{N}$ up to the order $\cO\left( 1/(\sqrt[K]{N})^{K+1} \right)$, while the remainder-term $\cR(N)$ captures any behavior of $\alpha(N)$ which does not have the form of this power series --- terms like $\log N$, or $N^N$, or various other types of asymptotics.

Now we define that the \textit{datapoints are ``nicely'' aligned along a $D$-dimensional hypersurface} if and only if in the limit $N\gg 1$ the asymptotics has the following form:
\begin{equation} \label{AsymptoticPropertiesDef}
\cR(N) = 0\,, \qquad c_K = 0\,, \qquad c_n = \left\{
\begin{array}{lcl}
c_D & & n=D\,, \\
0 & & n \neq D\,,
\end{array}
 \right. \qquad \forall n=0,\dots,K-1\,.
\end{equation}
This is actually a statement that $\alpha(N)$ has the asymptotics of the form \eqref{GoodAsymptotics}, while the single nonzero coefficient $c_D$ is proportional to the $D$-dimensional volume of the hypersurface,
\begin{equation}
c_D \equiv \frac{V_D}{(\sqrt[K]{V_K})^D}\,.
\end{equation}
Moreover, in this case one can solve \eqref{GoodAsymptotics} for $D$, which gives us the \textit{statistically defined dimension of the hypersurface},
\begin{equation} \label{StatisticalDimensionDef}
  D = K \left[ 1+ \lim_{N\to\infty} \frac{\log \alpha(N)}{\log N} \right]
  \,.
\end{equation}
Thus, if the asymptotics of $\alpha(N)$ is ``nice'' in the above sense, the quantity $D$ will be a positive integer smaller than $K$ (since $\log \alpha(N)$ is negative), and it can be explicitly computed, asymptotically for ever larger $N$.

Finally, if we have a bunch of available datasets (for example, $(N!)^6$, as we do in the main text), we can calculate the asymptotic form \eqref{GeneralAsymptotics} for each. Among all these, the one dataset that converges most efficiently toward \eqref{GoodAsymptotics} is called ``special'', and the corresponding value of $D$ calculated from \eqref{StatisticalDimensionDef} is called its \textit{dimension}. It defines a hypersurface $\cM$ as a strict subspace of the box of volume $V_K$, and is moreover of measure zero compared to the box, since
\begin{equation}
\frac{\meas(\cM)}{\meas(\text{box})} \equiv \lim_{N\to\infty} \frac{V_{total}}{V_K} = \lim_{N\to\infty} \alpha(N) = 0\,,
\end{equation}
according to the asymptotics \eqref{GoodAsymptotics}.

At the end, we note that we have implicitly also used the \textit{self-reinforcement} property of the dataset, in the sense that the limit $N\to\infty$ actually exists, i.e., that instead of the dataset of $N$ points we can use the dataset of $N+M$ points, such that the asymptotic properties \eqref{AsymptoticPropertiesDef} are maintained as we pass from one dataset to another. This is necessary for the limit $N\to\infty$ to actually make sense, in this context (see also \cite{Butterfield} for further discussion of the discrete-to-continuum limit).

In light of the asymptotic formulas \eqref{GeneralAsymptotics} and \eqref{AsymptoticPropertiesDef}, the three properties from the main text, namely (a) \textit{existence}, (b) \textit{self-reinforcement} and (c) \textit{dimensionality} mean in turn that (a) $c_K=\cR(N)=0$, (b) the limit $N\to\infty$ is well-defined, and (c) only one of the coefficients $c_n$ is different from zero. Finally, the fourth property, namely (d) \textit{topology}, can be determined from the overlapping $K$-dimensional cubes constructed around the datapoints. Specifically, in the limit $N\to \infty$ the total overlap volume $V_{overlap}$ must remain finite, precisely because $V_{total}$ falls to zero. This means that, despite the fact that each $K$-dimensional cube shrinks to zero, its overlap with neighboring cubes will remain finite. This property enables us to define {\em a basis of open sets} on our would-be manifold $\cM$ as the $D$-dimensional projections of all $K$-dimensional cubes (of volume $V_{cube}$), giving rise to a {\em topology on} $\cM$.

\subsection{Alternative technique for evaluating the critical parameter}

The algorithm for calculating $\alpha(N)$, while conceptually clear, turns out to be a bit inefficient for large datasets, due to the complexity of calculating the overlap volumes of neighboring cubes. The algorithm starts to choke already for $K=2$ and $N=100$ on a typical desktop computer. To remedy this, instead of using \eqref{AlphaDef} to calculate the parameter $\alpha$, one should instead split the whole box into a convenient grid of cells, and then count the cells which contain at least one datapoint. The total volume of the nonempty cells should then be similar to the total volume of the overlapping cubes from the original approach.

To implement this idea, we first choose the number of cells along each axis of the box to be
\begin{equation} \label{NumberOfCellsInTheBox}
\left\lfloor \sqrt[K]{N} \right\rfloor\,,
\end{equation}
so that the total number of cells is
\begin{equation}
\left\lfloor \sqrt[K]{N} \right\rfloor^K \leq N\,.
\end{equation}
This is the biggest integer smaller than $N$ such that the $K$-dimensional box can be divided into that many cells. The size of each cell in the $n$-th direction $x_n$ is then given as
\begin{equation}
\tilde\epsilon_n = \frac{x_n^{\text{max}} - x_n^{\text{min}}}{ \left\lfloor \sqrt[K]{N} \right\rfloor } \,,
\end{equation}
so that the total volume of each cell is
\begin{equation}
\tilde{V}_{\text{cell}} = \prod_{n=1}^K \tilde{\epsilon}_n = \frac{V_K}{\left\lfloor \sqrt[K]{N} \right\rfloor^K} \geq \frac{V_K}{N} \equiv V_{\text{cube}}\,,
\end{equation}
see \eqref{VolumeOfAsingleCube}. Given this arrangement, we can define a new parameter $\tilde\alpha$, in analogy to \eqref{VtotalDefDva}, as
\begin{equation} \label{TotalFilledVolumeNikola}
\tilde{V}_{\text{total}} \equiv p \tilde{V}_{\text{cell}} = \tilde{\alpha} V_K\,,
\end{equation}
where $p$ is the number of cells which contain at least one datapoint. We thus arrive at a new parameter,
\begin{equation}
\tilde{\alpha} = p \frac{\tilde{V}_{\text{cell}}}{V_K} = \frac{p}{\left\lfloor \sqrt[K]{N} \right\rfloor^K}\,.
\end{equation}
Note that we have used the ``floor'' function in \eqref{NumberOfCellsInTheBox} in order to avoid underestimating the volume \eqref{TotalFilledVolumeNikola}, and thus avoid underestimating $\tilde{\alpha}$ and consequently $D$.

Compared to \eqref{AlphaDef}, parameter $\tilde{\alpha}$ can be calculated much more efficiently, since it boils down to counting the number of nonempty cells $p$, which is way faster than evaluating all the overlapping volumes of cubes. Indeed, for the case $K=2$ and $N=10^4$, the algorithm takes around $10\, {\rm s}$ to evaluate $\tilde{\alpha}$ on the same hardware as before.

Intuitively, in the limit $N\to\infty$ one expects that $\tilde\alpha \approx \alpha$, at least in the case of ``nice'' alignment of datapoints (as defined by \eqref{AsymptoticPropertiesDef}). This establishes that in practical simulations we can calculate the statistical dimension $D$ of the dataset (see \eqref{StatisticalDimensionDef}) using $\tilde\alpha$ instead of $\alpha$, which is numerically much more efficient. Indeed, this is also confirmed by explicit numerical calculations on several different example datasets.

On the other hand, using the grid-like construction above obscures the notion of basis of open sets, since cells in the grid never overlap, and therefore one cannot use this construction to introduce the topology of the manifold. In this sense, while the grid-like construction of $\tilde{\alpha}$ is numerically more efficient, the overlapping-cubes construction of $\alpha$ gives us the information about topology and is thus conceptually more useful.

\end{document}